\documentclass[draftclsnofoot, onecolumn, 11pt]{IEEEtran}
\ifCLASSINFOpdf
\else
\fi

\usepackage{amssymb}
\usepackage{latexsym}
\usepackage{graphicx}
\usepackage{color}
\usepackage{verbatim}
\usepackage{multirow}
\usepackage{amsmath}

\usepackage{amsfonts}

\newcommand{\F}{\ensuremath{\mathbb F}}

\newcommand{\C}{\ensuremath{\mathbb C}}
\newcommand{\R}{\ensuremath{\mathbb R}}

\newcommand{\abu}{{\bf a}}

\newcommand{\xbu}{{\bf x}}
\newcommand{\ybu}{{\bf y}}
\newcommand{\wbu}{{\bf w}}
\newcommand{\zbu}{{\bf z}}

\newcommand{\Abu}{{\bf A}}

\newcommand{\Gbu}{{\bf G}}

\newcommand{\qed}{\hfill \ensuremath{\Box}}

\newtheorem{prop}{Proposition}
\newtheorem{df}{Definition}
\newtheorem{thr}{Theorem}

\newtheorem{rem}{Remark}
\newtheorem{cor}{Corollary}
\newtheorem{const}{Construction}

\newtheorem{const2}{Construction}
\numberwithin{const2}{const}

\begin{document}
%
\title{Deterministic Compressed Sensing Matrices from Additive Character Sequences}


\author{\IEEEauthorblockN{Nam Yul Yu} \\
\IEEEauthorblockA{Department of Electrical Engineering, Lakehead University \\
Email: nam.yu@lakeheadu.ca \\
}
September 30, 2010.
}


%


\maketitle

\begin{abstract}
Compressed sensing is a novel technique where one can
recover sparse signals from the
undersampled measurements.
In this correspondence, a $K \times N$ measurement matrix for compressed sensing
is deterministically constructed
via additive character sequences.
The Weil bound is then used to show that the matrix has asymptotically optimal coherence for $N=K^2$,
and to present a sufficient condition on the sparsity level for unique sparse recovery.
Also, the restricted isometry property (RIP) is statistically studied for the deterministic matrix.
Using additive character sequences with small alphabets,
the compressed sensing matrix can be efficiently implemented by linear feedback shift registers.
Numerical results
show that the deterministic compressed sensing matrix guarantees reliable matching pursuit recovery performance
for both noiseless and noisy measurements.
\end{abstract}


%

\section{Introduction}
Compressed sensing (or compressive sampling) is a novel and emerging technology
with a variety of applications in imaging, data compression, and communications.
In compressed sensing, one can recover sparse signals of high dimension
from few measurements that were believed to be incomplete.
Mathematically, measuring an $N$-dimensional signal $\xbu \in \R^N$
with a $K \times N$ measurement matrix $\Abu$ produces
a $K$-dimensional vector $\ybu = \Abu \xbu $ in compressed sensing, where $K<N$.
In recovering $\xbu$, it seems impossible to solve $K$ linear equations
with $N$ indeterminates by basic linear algebra.
However, imposing an additional requirement that $\xbu$ is $s$-sparse
or the number of nonzero entries in $\xbu$ is at most $s$,
one can recover $\xbu$ exactly with high probability
by solving the $l_1$-minimization problem,
which is computationally tractable.

Many research activities have been triggered on theory and practice of compressed sensing
since Donoho~\cite{Donoho:CS}, and Candes, Romberg, and Tao~\cite{CanRomTao:robust}\cite{CanTao:univ}
published their marvelous theoretical works.
The efforts revealed that
a measurement matrix $\Abu$ plays a crucial role in recovery of
$s$-sparse signals. 
In particular, Candes and Tao~\cite{CanTao:univ} presented the \emph{restricted isometry property (RIP)},
a sufficient condition for the matrix to guarantee sparse recovery. 
A \emph{random} matrix has been widely studied for the RIP,
where the entries are generated by a probability distribution such as the Gaussian or Bernoulli process,
or from randomly chosen partial Fourier ensembles.
Although a random matrix has many theoretical benefits~\cite{Rauhut:struct}, 
it has the drawbacks of high complexity, large storage, and low efficiency
in its practical implementation~\cite{Cald:large}.
As an alternative, 
we may consider 
a \emph{deterministic} matrix, where well known codes and sequences have been employed for the construction, 
e.g., chirp sequences~\cite{App:chirp}, Kerdock and Delsarte-Goethals codes~\cite{Cald:sub},
second order Reed-Muller codes~\cite{How:fast}, and dual BCH codes~\cite{Ailon:BCH}.
Other techniques for deterministic construction, based on finite fields, representations, and cyclic difference sets,
can also be found in~\cite{DeVore:det}$-$\cite{Yu:CDS}, respectively.
Although it is difficult to check the RIP 
and the theoretical recovery bounds are worse than that of a random matrix~\cite{Rauhut:struct},
the deterministic matrices guarantee reliable recovery performance in a statistical sense, 
allowing low cost implementation.

To enjoy the benefits of deterministic construction,
this correspondence presents how to construct a $K \times N$ measurement matrix for compressed sensing
via \emph{additive character sequences}.
We construct the matrix by employing additive character
sequences with small alphabets as its column vectors.
The Weil bound~\cite{Weil:BNT} is then used to show that the matrix has asymptotically optimal coherence for $N=K^2$, and
to present a sufficient condition on the sparsity level
for unique sparse recovery.
The RIP of the matrix is also analyzed through the eigenvalue statistics of the Gram matrices as in~\cite{App:chirp}. 
Using additive character sequences with small alphabets, 
the matrix can be efficiently implemented
by linear feedback shift registers.
Through numerical experiments,
we observe that the deterministic compressed sensing matrix
guarantees reliable and noise-resilient matching pursuit recovery performance 
for sparse signals.


\section{Preliminaries}
The following notations will be used throughout this correspondence.
\begin{enumerate}
\item[$-$] $\omega_p = e^{j \frac{ 2 \pi}{p}}$ is a primitive $p$-th root of unity, where $j = \sqrt{-1}$.
\item[$-$] $\F_q={\rm GF}(q)$ is the finite field with $q$ elements and 
$\F_q^*$ denotes the multiplicative group of $\F_q$.
\item[$-$] $\F_q [x]$ is the polynomial ring over $\F_q$, where each coefficient of $f(x) \in \F_q[x]$ is an element of $\F_q$.
\item[$-$] Let $p$ be prime, and $n$ and $m$ be positive integers with $m | n$.
A \emph{trace} function is a linear mapping from $\F_{p^n}$ onto $\F_{p^m}$ defined by
\[
{\rm Tr}_m ^n (x) = \sum_{i=0} ^{n/m -1} x^{p^{mi}}, \quad x \in \F_{p^n}
\]
where the addition is computed modulo $p$.
\end{enumerate}

\subsection{Additive characters}

Let $p$ be prime and $m$ a positive integer.
We define an \emph{additive character}~\cite{Lidl:FF} of $\F_{p^m}$ as
\begin{equation}\label{eq:add_tr}
\chi(x) = \exp \left(j \frac{ 2 \pi  {\rm Tr}_1 ^m ( x)}{p}  \right), \quad x \in \F_{p^m}
\end{equation}
where $\chi(x+y) = \chi(x) \chi(y)$ for $x, y \in \F_{p^m}$.
The Weil bound~\cite{Weil:BNT} gives an upper bound on the magnitude of additive character sums.
We introduce the bound as described in~\cite{Lidl:FF}.

\vspace{0.075in}
\begin{prop}~\cite{Lidl:FF}\label{prop:Weil}
Let $f(x)\in \F_{p^m}[x]$ be a polynomial of degree $r \geq 1$ with $\gcd(r, p^m)=1$.
Let $\chi$ be the additive character of $\F_{p^m}$ defined in (\ref{eq:add_tr}).
Then,
\begin{equation}\label{eq:Weil_add}
\left| \sum_{x\in \F_{p^m}} \chi (f(x)) \right| \leq \left(r -1 \right) \sqrt{p^m}.
\end{equation}
\end{prop}
\vspace{0.075in}

From (\ref{eq:Weil_add}), $\sum_{x\in \F_{p^m}} \chi (x) = 0$ is obvious.
In the Weil bound, the condition of $\gcd(r, p^m)=1$
can be replaced by the weaker one that the polynomial $f$ should not be of
the form  $h(x)^p+h(x)+e$ for any polynomial $h(x) \in \F_{p^m} [x]$
and any $e \in \F_{p^m}$~\cite{Lidl:FF}.

\subsection{Restricted isometry property}

The restricted isometry property (RIP)~\cite{CanTao:univ} presents a sufficient condition for a measurement matrix $\Abu$
to guarantee unique sparse recovery.
\vspace{0.075in}
\begin{df}\label{def:rip}
The restricted isometry constant $\delta_s$ of a $K \times N$ matrix $\Abu$ is defined
as the smallest number such that
\begin{equation*}\label{eq:rip}
(1-\delta_s) || \xbu ||_{l_2} ^2 \leq || \Abu \xbu ||_{l_2} ^2 \leq (1+\delta_s) || \xbu ||_{l_2} ^2
\end{equation*}
holds for all $s$-sparse vectors $\xbu \in \R^N$,
where $||\xbu ||_{l_2} ^2 = \sum_{n=0} ^{N-1} |x_n|^2$ with
$\xbu = (x_0, \cdots, x_{N-1}) \in \R^N$.
\end{df}
\vspace{0.075in}
We say that $\Abu$ obeys the RIP of order $s$ if $\delta_s$ is reasonably small, not close to $1$.
In fact, the RIP requires that
all subsets of $s$ columns taken from the measurement matrix should be \emph{nearly orthogonal}~\cite{CanWak:intro}.
Indeed, Candes~\cite{Can:rip} asserted that if $\delta_{2s} <1$, a unique $s$-sparse solution
is guaranteed by $l_0$-minimization, which is however a hard combinatorial problem.

A tractable approach for sparse recovery is to solve the $l_1$-minimization~\cite{CanRomTao:robust}, i.e.,
to find a solution of
$\min_{\widetilde{\xbu} \in \R^N} || \widetilde{\xbu} ||_{l_1}$ subject to $\Abu \widetilde{\xbu} = \ybu$,
where $||\widetilde{\xbu} ||_{l_1} = \sum_{i=0} ^{N-1} |\widetilde{x}_i|$.
In addition,
greedy algorithms~\cite{Tropp:greed} have been also proposed
for sparse signal recovery,
including matching pursuit (MP)~\cite{Mallat:mp}, orthogonal matching pursuit (OMP)~\cite{Tropp:omp},
and CoSaMP~\cite{Needell:cosamp}.
In particular, if a measurement matrix is deterministic, 
we may exploit its structure to develop
a reconstruction algorithm for sparse signal recovery~\cite{App:chirp}\cite{How:fast}\cite{Yu:CDS},
providing fast processing and low complexity.

\subsection{Coherence and redundancy}
In compressed sensing, a $K \times N$ deterministic matrix $\Abu$ is associated with
two geometric quantities, \emph{coherence}
and \emph{redundancy}~\cite{Tropp:gap}.
The coherence $\mu$ is defined by
\begin{equation*}
\mu = \max_{0 \leq l \neq m \leq N-1} \left| \abu_l ^H \cdot \abu_m \right|
\end{equation*}
where $\abu_*$ denotes a column vector of $\Abu$ with $|| \abu_* ||_{l_2} = 1$,
and $\abu_* ^H$ is its conjugate transpose.
In fact, the coherence is a measure of mutual orthogonality among the columns,
and the small coherence is desired for good sparse recovery~\cite{Rauhut:struct}.
In general, the coherence is lower bounded by
\begin{equation*}\label{eq:welch}
\mu \geq \sqrt{\frac{N-K}{K(N-1)}}
\end{equation*}
which is called the \emph{Welch bound}~\cite{Welch:low}.

The redundancy, on the other hand, is defined as $\rho = || \Abu ||^2$,
where $|| \cdot ||$ denotes the spectral norm of $\Abu$,
or the largest singular value of $\Abu$.
We have $\rho \geq N/K$, where the equality holds if and only if $\Abu$
is a \emph{tight} frame.
For unique sparse recovery, it is desired that $\Abu$ should be a tight frame with the smallest redundancy~\cite{Cald:LASSO}.

\section{Compressed Sensing Matrices from Additive Character Sequences}


\subsection{Construction}

\begin{const}\label{cst:mat_add}
Let $p$ be an odd prime, and $m$ and $h$ be positive integers
where $h>1$.
Let $K=p^m$ and $N=K^h=p^{mh}$. 
Set a column index to $n = \sum_{i=1} ^{h} u_i K^{i-1}$
where $u_i = \left \lfloor \frac{n}{K^{i-1}} \right \rfloor \mod{K}$.
For each $i$, $1 \leq i \leq h$, let
\begin{equation}\label{eq:b_u}
b_i = \left \{ \begin{array}{ll} 0, & \quad \mbox{if } u_i = 0, \\
\alpha^{u_i-1}, & \quad \mbox{if } 1 \leq u_i \leq p^m-1  \end{array} \right.
\end{equation}
where $b_i \in \F_{p^m}$ and $\alpha$ is a primitive element in $\F_{p^m}$.
For a positive integer $d \geq h$, let $r_1, r_2, \cdots, r_h$ be $h$ distinct integers such that
$1=r_1 < r_2 < \cdots < r_h=d$ and $\gcd(r_i, p^m) = 1$ for each $i$, $1 \leq i \leq h$.
Then, we construct a $K \times N$ compressed sensing matrix $\Abu$ where each entry is given by
\begin{equation}\label{eq:phi_add}
a_{k, n} = \left \{ \begin{array}{ll} \frac{1}{\sqrt{K}}, & \quad \mbox{if } k=0, \\
\frac{1}{\sqrt{K}} \omega_p ^{{\rm Tr}_1 ^m \left( \sum_{i=1} ^h b_i \alpha^{r_i(k-1)} \right)}, & \quad \mbox{if } 1 \leq k \leq K-1 \end{array} \right.
\end{equation}
where $ 0 \leq n \leq N-1$.
\end{const}
\vspace{0.075in}

In Construction~\ref{cst:mat_add},
if $p=2$ and $r_i$'s are successive odd integers,
then each column vector of $\Abu$ 
is equivalent to a codeword of the dual of the extended binary BCH code,
which has been studied in~\cite{Ailon:BCH} for compressed sensing.
In~\cite{Xu:trig}, Xu also presented a similar construction by
defining an additive character with large alphabet as $\chi(x) = e^{j 2 \pi x/p}$ where $p=K$,
which is a \emph{generalization} of chirp sensing codes~\cite{App:chirp}. 


Using the Weil bound on additive character sums, we determine the coherence of $\Abu$.

\vspace{0.075in}
\begin{thr}\label{th:coh_add}
In the $K \times N$ matrix $\Abu$ from Construction~\ref{cst:mat_add},
the coherence is given by
\begin{equation}\label{eq:coh_add}
\mu = \max_{0 \leq n_1 \neq n_2 \leq N-1} \left| \abu_{n_1} ^H \cdot \abu_{n_2} \right| =
\frac{d-1}{\sqrt{K}}
\end{equation}
where if $d=2$, the coherence is
asymptotically optimal, achieving the equality of the Welch bound.
\end{thr}
\vspace{0.075in}

\noindent \textit{Proof.}
Consider the column indices of $n_1 = \sum_{i=1} ^{h} u_i K^{i-1}$ and $n_2 = \sum_{i=1} ^{h} u_i ' K^{i-1}$, where $n_1 \neq n_2$.
According to (\ref{eq:b_u}), let $b_i = 0$ or $\alpha^{u_i-1}$, and $b_i ' = 0$ or $\alpha^{u_i '-1}$, respectively.
Similarly, from (\ref{eq:phi_add}), let $x=0$ if $k=0$, or $x = \alpha^{k-1}$ otherwise.
Then, the inner product of a pair of columns in $\Abu$ is given by
\begin{equation}\label{eq:add_f}
\begin{split}
\left| \abu_{n_1} ^H \cdot \abu_{n_2} \right|
 = \frac{1}{K} \left| \sum_{x \in \F_{p^m}} \omega_p ^{{\rm Tr}_1 ^m \left(\sum_{i=1} ^h (b_i ' - b_i) x^{r_i} \right)} \right|
 = \frac{1}{K} \left| \sum_{x \in \F_{p^m}} \chi\left( \sum_{i=1} ^h (b_i ' - b_i ) x^{r_i}  \right) \right| .
\end{split}
\end{equation}
In (\ref{eq:add_f}), if $n_1 \neq n_2$, then $f(x) = \sum_{i=1} ^h (b_i ' - b_i) x^{r_i}$
is a nonzero polynomial in $\F_{p^m}[x]$,
as there exists at least a pair of $(b_i, b_i ')$ where $b_i \neq b_i '$.
Since $\gcd(r_i, p^m)=1$ for any $r_i$,
the Weil bound in Proposition~\ref{prop:Weil} produces
\[
\left| \abu_{n_1} ^H \cdot \abu_{n_2} \right| \leq \frac{(d-1)\sqrt{K}}{K} =  \frac{d-1}{\sqrt{K}}
\]
from which the coherence $\mu$ in (\ref{eq:coh_add}) is immediate.
For given $K$ and $N$,
the equality of the Welch bound is computed by
$ \sqrt{\frac{N-K}{K(N-1)}} = \sqrt{\frac{K^h-K}{K^{h+1}-K}} \approx \frac{1}{\sqrt{K}}$.
Therefore, the coherence
asymptotically achieves the equality of the Welch bound if $d=2$.
\qed
\vspace{0.075in}


\vspace{0.075in}
\begin{thr}\label{th:s_cds}
In the matrix $\Abu$ from Construction~\ref{cst:mat_add},
unique $s$-sparse signal recovery is guaranteed by $l_1$-minimization or greedy algorithms if
\begin{equation}\label{eq:s_cds}
s< \frac{1}{2} \left(\frac{\sqrt{K}}{d-1} + 1 \right).
\end{equation}
\end{thr}
\vspace{0.075in}
\noindent \textit{Proof.}
The upper bound on the sparsity level 
is straightforward from the coherence $\mu = \frac{d-1}{\sqrt{K}}$
and the Tropp's sufficient condition~\cite{Tropp:greed} of $s < \frac{1}{2}(\mu^{-1}+1)$ for unique sparse recovery.
\qed

\vspace{0.075in}
\begin{rem}\label{rm:sparsity}
In Construction~\ref{cst:mat_add},
if $d=h$, then
$\log \left(\frac{N}{K} \right) = (h-1) \log K = (d-1) \log K$
with $N = K^h$.
Thus, $\frac{1}{d-1} = \frac{\log K}{\log (N/K)}$ and
from (\ref{eq:s_cds}), we have the sparsity bound of
\[
s < \frac{1}{2} \left(\frac{\sqrt{K} \log K}{\log (N/K)} +1 \right)
\]
for unique sparse recovery.
Therefore, we obtain the sparsity bound of $s \leq C \sqrt{K} \log K/ \log (N/K) $
from Construction~\ref{cst:mat_add},
which is known to be the largest for deterministic construction~\cite{DeVore:det}.
\end{rem}

\vspace{0.075in}
\begin{thr}\label{th:red_add}
In Construction~\ref{cst:mat_add},
$\Abu$ is a tight frame with redundancy $\rho = N/K$.
\end{thr}
\vspace{0.075in}

\noindent \textit{Proof.}
In $\Abu$, consider a set of $K$ column indices of $n = \sum_{i=1} ^{h} u_i K^{i-1}$
where $u_1$ is varying for $0 \leq u_1 \leq K-1$,
while $u_2, u_3, \cdots, $ and $u_h$ are fixed.
Accordingly, note from (\ref{eq:b_u}) that $b_1$ runs through $\F_{p^m}$,
while $b_2, b_3, \cdots, $ and $b_h$ are fixed in $\F_{p^m}$.
Then, the set of columns forms a $K \times K$ submatrix $\sigma_t$,
where $t=u_2 + u_3 K+ \cdots + u_h K^{h-2}$, $0 \leq t \leq K^{h-1}-1$.
In fact, $\sigma_t$ is a set of $K$ orthonormal bases and $\Abu$ is a concatenation of the $K^{h-1}$ sets
for $0 \leq t \leq K^{h-1}-1$.

Let $\wbu_{k_1}$ and $\wbu_{k_2}$ be
a pair of distinct row vectors in $\sigma_t$, where $0 \leq k_1 \neq k_2 \leq K-1$.
For $i=1$ and $2$, let $x_i = 0$ if $k_i = 0$, and $x_i = \alpha^{k_i-1}$ if $k_i \geq 1$.
Then, the inner product of $\wbu_{k_1}$ and $\wbu_{k_2}$ is given by
\begin{equation*}
\begin{split}
\left| \wbu_{k_1}  \cdot \wbu_{k_2} ^H \right|
& = \frac{1}{K} \left| \sum_{b_1 \in \F_{p^m}} \omega_p ^{{\rm Tr}_1 ^m \left(\sum_{i=1} ^h b_i (x_1 ^{r_i} - x_2 ^{r_i}) \right)} \right| \\
& = \frac{1}{K} \left| \omega_p ^{{\rm Tr}_1 ^m \left(\sum_{i=2} ^h b_i (x_1 ^{r_i} - x_2 ^{r_i}) \right)} \right| \cdot
\left| \sum_{b_1 \in \F_{p^m}} \chi\left( (x_1 ^{r_1} -  x_2 ^{r_1}) b_1  \right) \right| \\
& = 0
\end{split}
\end{equation*}
where $x_1 \neq x_2$.
Finally, the mutual orthogonality of a pair of distinct rows in $\Abu$ is obtained
if the submatrix $\sigma_t$ is concatenated for all $t$.
Similar to the proof of Lemma 6 in~\cite{Cald:large}, $\Abu$ is a tight frame with redundancy $\rho = N/K$
from the mutual orthogonality
and $|a_{k, n} |= \frac{1}{\sqrt{K}}$.
\qed
\vspace{0.075in}

Theorem~\ref{th:coh_add} proved that
the compressed sensing matrix $\Abu$ in Construction~\ref{cst:mat_add}
has optimal coherence if $d=2$.
Also, Theorem~\ref{th:red_add} showed that $\Abu$ is a tight frame
with optimal redundancy $\rho=N/K$ for any choice of $d$.
Taking $d=h=2$, and $r_1 = 1$ and $ r_2 = 2$,
we consider a special matrix
of Construction~\ref{cst:mat_add}.

\vspace{0.075in}
\begin{const2}\label{cst:mat_add2}
Let $p$ be an odd prime and $m$ be a positive integer.
Let $K=p^m$ and $N=K^2=p^{2m}$. 
Set a column index to $n = u_1 + u_2 K$
where $u_1 = n \pmod{K}$ and $u_2 = \left \lfloor \frac{n}{K} \right \rfloor$.
For $i=1$ and $2$, define $b_i$ by (\ref{eq:b_u}).
Then, we construct a $K \times N$ compressed sensing matrix $\Abu$ where each entry is given by
\begin{equation*}\label{eq:phi_add2}
a_{k, n} = \left \{ \begin{array}{ll} \frac{1}{\sqrt{K}}, & \quad \mbox{if } k=0, \\
\frac{1}{\sqrt{K}} \omega_p ^{{\rm Tr}_1 ^m \left( b_1 \alpha^{(k-1)} + b_2 \alpha^{2(k-1)} \right)}, & \quad \mbox{if } 1 \leq k \leq K-1 \end{array} \right.
\end{equation*}
where $ 0 \leq n \leq N-1$.
\end{const2}
\vspace{0.075in}

From Theorems~\ref{th:coh_add}, \ref{th:s_cds}, and \ref{th:red_add},
the following properties of $\Abu$ in Construction~\ref{cst:mat_add2} are straightforward.

\vspace{0.075in}
\begin{cor}\label{co:add2}
The coherence of
$\Abu$ in Construction~\ref{cst:mat_add2} is
$ \mu = \frac{1}{\sqrt{K}}$.
Then, the matrix $\Abu$ guarantees unique $s$-sparse signal recovery by $l_1$-minimization or greedy algorithms if
$s< \frac{1}{2} \left(\sqrt{K} + 1 \right)$.
Also, 
$\Abu$ is a tight frame with redundancy $K$.
With the optimal coherence and redundancy, the $K \times K^2$ compressed sensing matrix $\Abu$ 
is an \emph{ideal} deterministic construction for unique sparse recovery~\cite{Cald:LASSO}.
\end{cor}


\subsection{RIP analysis}

\begin{figure}[!t]
\centering
\includegraphics[width=0.75\textwidth, angle=0]{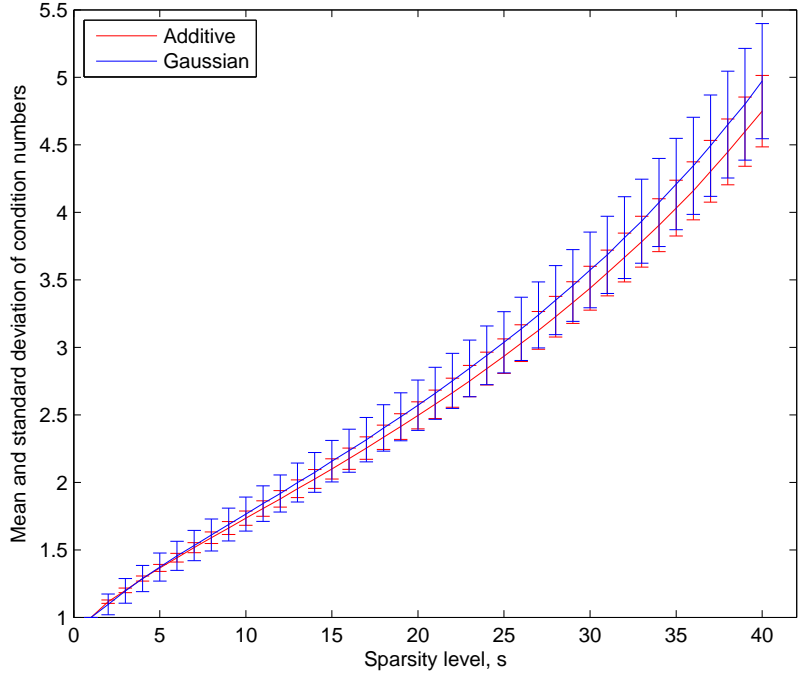}
\caption{Condition number statistics of additive character and Gaussian random sensing matrices. ($K=81, N=6561, p=3$)}
\label{fig:eigen}
\end{figure}

In Construction~\ref{cst:mat_add2}, we choose $p=3$ and $m=4$,
which produces the compressed sensing matrix $\Abu$ with $K=81$ and $N=6561$. 
For statistical analysis, we measured the condition number, or the ratio
of the largest singular value of a matrix to the smallest.
Figure~\ref{fig:eigen} displays the means and standard deviations of the condition numbers
of $\Abu_s$ and $\Gbu_s $, respectively, where %
$\Abu_s$ is a submatrix of $s$ columns randomly chosen from the additive character sensing matrix $\Abu$, 
while $\Gbu_s$ is a $K \times s$ Gaussian random matrix with $K=81$.
The statistics were measured over total $10, 000$ condition numbers,
where each matrix is newly chosen at each instance.
Each entry of the Gaussian matrix $\Gbu_s$ is independently sampled from the Gaussian distribution
of zero mean and variance $\frac{1}{K}$, and each column vector is then normalized such that
it has unit $l_2$-norm.

Note that the singular values of $\Abu_s$ are the square roots of the eigenvalues of the Gram matrix $\Abu_s ^H \Abu_s$.
Since the RIP requires that the Gram matrix should have all the eigenvalues
in an interval $[1-\delta_s, 1+\delta_s]$ with reasonably small $\delta_s$~\cite{Rauhut:struct},
the condition numbers should be as small as possible for unique sparse recovery.
From this point of view,
we observe from Figure~\ref{fig:eigen} that our additive character sensing matrix
shows better statistics of condition numbers than the Gaussian matrix.
This convinces us that $\Abu$ in Construction~\ref{cst:mat_add2}
is suitable for compressed sensing
in a statistical sense.

\subsection{Implementation}


In Construction~\ref{cst:mat_add},
excluding the first element of $\frac{1}{\sqrt{K}}$ ,
each column of $\Abu$
is a \emph{pseudo-random sequence}
where each element is represented as a combination of trace functions which is modulated by an exponential function.
Precisely, the pseudo-random sequence is 
$c_{b_1, b_2, \cdots, b_h} (k) = 
\sum_{i=1} ^h {\rm Tr}_1 ^m \left(b_i \alpha^{r_i k} \right), \ k=0,1, \cdots, p^m-2 $.
Since a sequence of a trace function 
is generated by a linear feedback shift register (LFSR)~\cite{GolGong:SD},
$c_{b_1, b_2, \cdots, b_h} (k)$ is generated by a combination of $h$ different LFSRs
where each LFSR has at most $m$ registers.
Generating each column with LFSRs,
we can efficiently implement the sensing matrix $\Abu$ with low complexity.
For more details on a trace function and its LFSR implementation, see \cite{GolGong:SD}.

\begin{figure}
\centering
\input{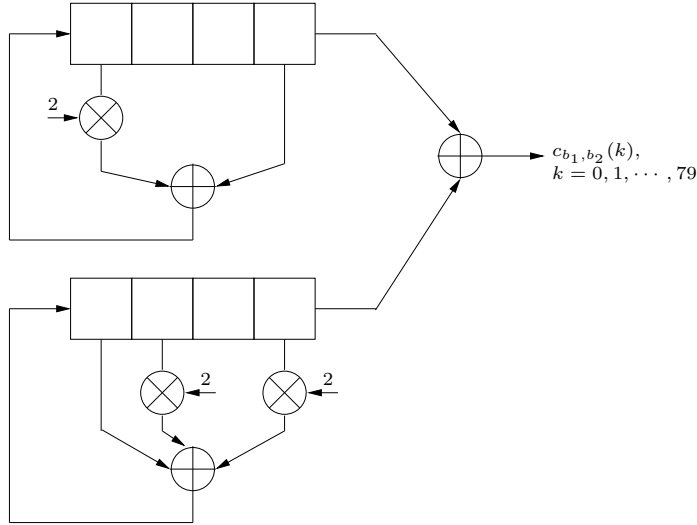}
\caption{The LFSR implementation of $c_{b_1, b_2}(k)$ for $\Abu$ in Construction~\ref{cst:mat_add2}, where $p=3$ and $m=4$.}
\label{fig:lfsr_add}
\end{figure}

As an example, Figure~\ref{fig:lfsr_add} illustrates an LFSR implementation 
generating a sequence
$c_{b_1, b_2}(k) = {\rm Tr}_1 ^m \left(  b_1 \alpha^{k} \right) + {\rm Tr}_1 ^m \left(  b_2 \alpha^{2k} \right), \ k =0, 1, \cdots, p^m-2$,
for the matrix $\Abu$ in Construction~\ref{cst:mat_add2}.
In the example, we take $p=3$ and $m=4$,
and define the finite field $\F_{3^4}$ by a primitive polynomial $g_1(x) = x^4 + x^3 +2$
that has the roots of a primitive element $\alpha$ and its conjugates.
Then, $g_1(x)$ specifies a feedback connection of the upper LFSR that generates a ternary sequence of
${\rm Tr}_1 ^4 (b_1 \alpha^k)$.
The lower LFSR, on the other hand, has a feedback connection specified by $g_2(x) = x^4+2 x^3+x^2+1$
that has the roots of $\alpha^2$ and its conjugates,
generating a ternary sequence of ${\rm Tr}_1 ^4 (b_2 \alpha^{2k})$. 
In the structure, note that each register can take a value of $0, 1$, or $2$,
and the addition and multiplication are computed modulo $3$.
Finally, the sequences of $c_{b_1, b_2}(k)$ are generated by the LFSR structure for
every possible pairs of initial states corresponding to $b_1, \ b_2 \in \F_{3^4}$.
As there exist total $3^{8}$ initial state pairs, the corresponding sequences  $c_{b_1, b_2}(k)$
make $N=3^{8}$ columns for $\Abu$.

\section{Recovery performance}

\subsection{Recovery from noiseless data}

\begin{figure}[!t]
\centering
\includegraphics[width=0.75\textwidth, angle=0]{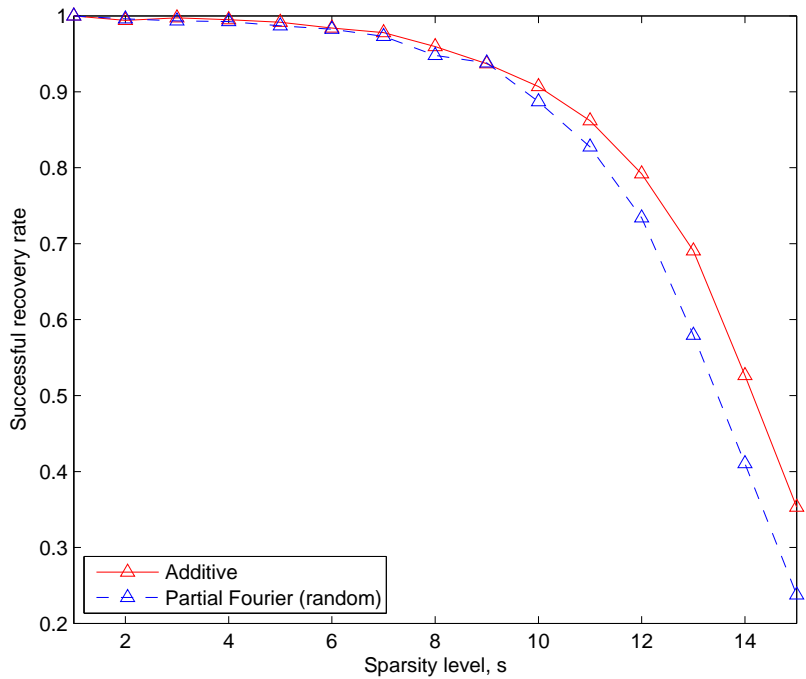}
\caption{Successful recovery rates for additive character and partial Fourier sensing matrices, where $K=81$ and $ N=6561$.}
\label{fig:succ_add}
\end{figure}

Figure~\ref{fig:succ_add} shows numerical results of successful recovery rates of $s$-sparse signals
measured by a $81 \times 6561$ compressed sensing matrix $\Abu$ in Construction~\ref{cst:mat_add2},
where total $2000$ sample vectors were tested for each sparsity level.
For comparison, the figure also shows the rates for randomly chosen partial Fourier matrices of the same dimension,
where we chose a new matrix at each instance
of an $s$-sparse signal,
in order to obtain the average rate.
Each nonzero entry of an $s$-sparse signal $\xbu$ is independently sampled from the normal distribution
with zero mean and variance $1$, where its position is chosen uniformly at random.
For both sensing matrices, the matching pursuit recovery with maximum iteration of $100$ was applied for the reconstruction of sparse signals.
A success is declared in the reconstruction if the squared error is reasonably small for the estimate $\widehat{\xbu}$, i.e.,
$|| \xbu - \widehat{\xbu} ||_{l_2} ^2 < 10^{-4}$.

In the experiment, we observed that if $s\leq 4$, more than $99 \%$ of $s$-sparse signals
are successfully recovered for the matrix $\Abu$, which verifies the sufficient condition in Corollary~\ref{co:add2}.
Furthermore, the figure reveals that our sensing matrix 
has fairly good recovery performance as the sparsity level increases.
For instance, more than $95 \%$ successful recovery rates are observed for $s \leq 7$,
which implies that the sufficient condition is a bit pessimistic.
The sensing matrix also shows better recovery performance than randomly chosen partial Fourier matrices with matching pursuit recovery.
We made a similar observation from additive character and partial Fourier compressed
sensing matrices with $K=49$ and $N=2401$.

In Figure~\ref{fig:succ_add},
each element of $\Abu$
takes $\frac{1}{9}$, $\frac{1}{9} e^{j\frac{2 \pi}{3}}$, or $\frac{1}{9} e^{j\frac{4 \pi}{3}}$, 
while the partial Fourier matrix has the element of $\frac{1}{9} e^{j\frac{2 \pi}{N}}$ where $N=6561$.
Therefore, the compressed sensing matrix from additive character sequences with small alphabets
has low implementation complexity as well as good recovery performance.


\subsection{Recovery from noisy data}

\begin{figure}[!t]
\centering
\includegraphics[width=0.75\textwidth, angle=0]{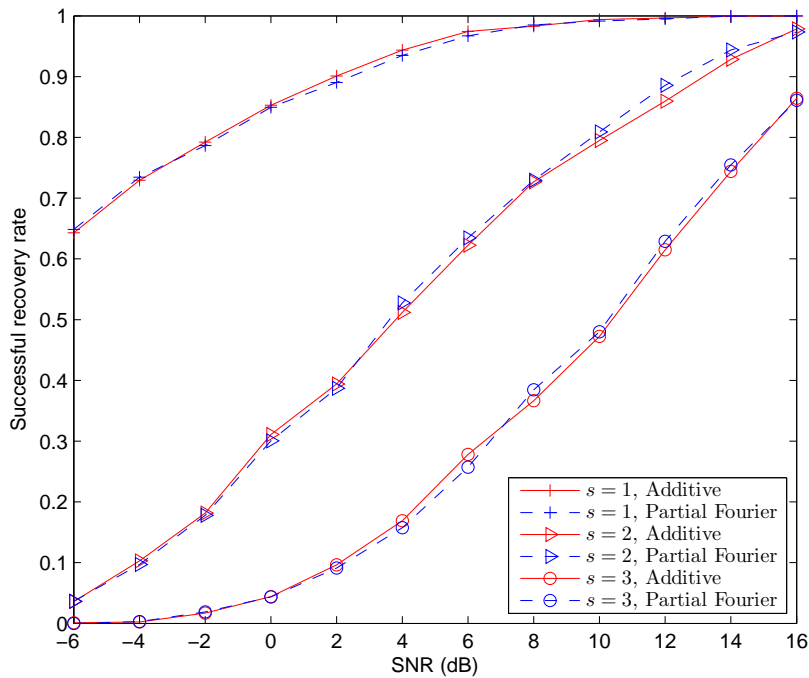}
\caption{Successful recovery rates for additive character and partial Fourier sensing matrices in the presence of noise, where $K=81$ and $N=6561$.}
\label{fig:succ_n}
\end{figure}

In practice, a measured signal $\ybu$ contains measurement noise, i.e.,
$\ybu = \Abu \xbu + \zbu$, where $\zbu \in \C^K$ denotes a $K$-dimensional complex vector of noise.
Thus, a compressed sensing matrix must be robust to measurement noise
by providing stable and noise resilient recovery.
Figure~\ref{fig:succ_n} displays the matching pursuit recovery performance of our sensing matrix
in the presence of noise.
The experiment parameters and the sparse signal generation are identical to those of noiseless case.
In the figure, $\xbu$ is $s$-spare for $s=1, 2,3$, and
signal-to-noise ratio (SNR) is defined by
${\rm SNR} 
= \frac{||\Abu \xbu ||_{l_2} ^2}{K \sigma_z^2}$,
where each element of $\zbu$ is an independent and identically distributed (i.i.d.)
complex Gaussian random process with zero mean and variance $\sigma_z ^2$.
In noisy recovery, a success is declared if 
$|| \xbu - \widehat{\xbu} ||_{l_2} ^2 < 10^{-2}$ after $100$ iterations.
From Figure~\ref{fig:succ_n}, we observe that the recovery performance 
is stable and robust against noise corruption at sufficiently high SNR,
which is
similar to that of randomly chosen partial Fourier matrices.

\section{Conclusion}
This correspondence has presented how to deterministically construct a $K \times N$ measurement matrix for compressed sensing
via additive character sequences.
We presented a sufficient condition on the sparsity level of the matrix for unique sparse recovery.
We also showed that the deterministic matrix with $N=K^2$
is ideal, achieving the optimal coherence and redundancy.
Furthermore, the RIP of the matrix has been statistically analyzed,
where we observed that it has better eigenvalue statistics than Gaussian random matrices.
The compressed sensing matrix from additive character sequences
can be efficiently implemented using LFSR structure.
Through numerical experiments,
the matching pursuit recovery of
sparse signals showed reliable and noise resilient performance
for the compressed sensing matrix. 







%

\end{document}